\documentclass[11pt]{article}
\usepackage{geometry}
\usepackage[utf8]{inputenc}
\usepackage{amsmath,amsfonts,amssymb}
\usepackage{breqn}
\usepackage{graphicx}
\usepackage{xfrac}
\usepackage{authblk}
\usepackage{enumitem}
\usepackage[colorlinks,citecolor=blue,urlcolor=black]{hyperref}
\usepackage[capitalise]{cleveref}
\usepackage[numbers, compress]{natbib}
\usepackage[font=small,labelfont=bf]{caption}
\usepackage{multicol}
\usepackage{subcaption}
\usepackage{titlesec}
\usepackage{ragged2e}
\usepackage{floatrow}
\usepackage{booktabs}
\usepackage{varwidth}
\usepackage{wrapfig}

\newsavebox\tmpbox

\newfloatcommand{capbtabbox}{table}[][\FBwidth]

\newenvironment{Figure}
  {\par\medskip\noindent\minipage{\linewidth}}
  {\endminipage\par\medskip}

\title{Rapidity gap distribution in diffractive dissociation: predictions for future electron-ion colliders}
\author[(1)]{Anh Dung Le}
\affil[(1)]{\em CPHT, CNRS, {\'E}cole polytechnique, IP Paris, F-91128 Palaiseau, France}
\date{\today}

\begin{document}
\maketitle

\begin{abstract}
	We present predictions for the distribution of rapidity gaps in realistic kinematics of future electron-ion colliders, based on numerical solutions of the original Kovchegov-Levin equation and of its next-to-leading extension taking into account the running of the strong coupling. We find that for the rapidities we have considered, the fixed and the running coupling equations lead to different distributions, rather insensitive to the chosen prescription in the running coupling case. The obtained distributions for the fixed coupling framework exhibit a shape characteristic of a recently proposed partonic picture of diffractive dissociation already at rapidities accessible at future electron-ion colliders. The modification of this shape in the running coupling case can also be understood qualitatively from that picture. Our results confirm the relevance of measurements of such observables for the microscopic understanding of diffractive dissociation in the framework of quantum chromodynamics.  
\end{abstract}

\section{\large INTRODUCTION}
\label{sec:introduction}

The observation of diffractive events accounting for about 10\% of all events in deeply inelastic electron-proton collisions at DESY HERA \cite{h1.1995,zeus.1995} was a striking experimental discovery. By definition, a diffractive event has a large rapidity gap in the final state~\cite{bjorken.1994}, namely a large angular sector in detector in which no particle is measured, which is interpreted as the signature of color singlet exchange. As the high rate of such events can be interpreted as a smoking gun for the onset of high parton density effects~\cite{gbw.1999,gbw.1998,gbw.2001,munier.etal.2001}, and diffractive events can give insights into the spatial distribution of the gluonic content of hadrons (for a review, see Ref.~\cite{Mantysaari.2020}), their detailed investigation is among main goals at future electron-ion colliders~\cite{EIC.2016,LHeC.2012}.

Of particular interest for diffractive deep-inelastic scattering (diffractive DIS) processes is diffractive vector meson ($V$) production, $\gamma^*h \to Vh'$, and diffractive dissociation, $\gamma^* h \to Xh'$, where the virtual photon mediating the interaction is dissociated into an inclusive set of particles ($X$) in the final state ($h'$ is either the hadron $h$ or an excited state of $h$ with the same quantum numbers). In recent years, the knowledge on the former has been advanced greatly by numerous studies (see e.g., Refs.~\cite{Mantysaari.2020,Deak.etal.2021} and references therein). The diffractive dissociation has also been investigated for a long time (see Ref.~\cite{kovchegov.levin.2012} for a review). Nevertheless, recent developments in this topic are quite limited, especially in understanding the diffraction at a microscopic level. The partonic content of the hadrons has signatures in the final state, such as the size of the rapidity gap. Therefore, the rapidity gap distribution is an important observable, as it can provide indications on the detailed microscopic mechanism of the diffraction.

In order to address that diffractive observable, an elegant formulation was established in the framework of quantum chromodynamics (QCD)~\cite{kovchegov.levin.2000,levin.lublinsky.2001,levin.lublinsky.2002a,levin.lublinsky.2002b,kovner.wiedemann.2001,hentchinski.weigert.schafer.2006,kovner.lublinsky.weigert.2006,hatta.etal.2006,kovchegov.2012}, which provides detailed predictions for diffractive cross sections of the scattering of a quark-antiquark ($q\bar{q}$) dipole with a nucleus, in the form of nonlinear evolution equations. It is referred to as the Kovchegov-Levin formulation, which relies on the color dipole picture \cite{gbw.1999,gbw.1998,gbw.2001,nikolaev.zakharov.1991a,nikolaev.zakharov.1991b,nikolaev.zakharov.1992}. The first investigation on the rapidity gap distribution in the dipole-nucleus scattering was already presented in Ref.~\cite{kovchegov.levin.2000} based on the analytical solution of a simplified version of the Kovchegov-Levin equation at leading order. This equation was then studied numerically in Ref.~\cite{levin.lublinsky.2001,levin.lublinsky.2002a,levin.lublinsky.2002b}, with a short discussion on the rapidity gap distribution in the nuclear scattering of the dipole~\cite{levin.lublinsky.2002a}. The diffractive onium-nucleus cross section at a fixed rapidity gap, which is directly related to the rapidity gap distribution by a normalization to the total cross section (see \cref{eq:photon_gap_distrib} below), has also been investigated recently in double log approximation~\cite{Contreras.etal.2018}, based on the same formalism. In a slightly different approach, there has been an attempt recently in deriving the rapidity gap distribution in the diffractive dissociation of small dipoles off nuclei at an asymptotic high energy based on the color dipole model~\cite{munier.mueller.2018a,munier.mueller.2018b}. Additionally, there were also different analytical and numerical model-dependent analyses on the diffractive mass spectrum~\cite{EIC.2016,nikolaev.zakharov.1992}, which is related through a simple identity to the rapidity gap distribution. 

In this work, we mainly study the rapidity gap distribution in diffractive deep-inelastic virtual photon-nucleus scattering. The investigation is based on the aforementioned QCD dipole model of DIS and numerical solutions to nonlinear evolution equations in both fixed coupling~\cite{kovchegov.levin.2000,kovner.wiedemann.2001,hentchinski.weigert.schafer.2006,kovner.lublinsky.weigert.2006,hatta.etal.2006,balitsky.1996,kovchegov.1999} and running coupling scenarios~\cite{kovchegov.2012,albacete.etal.2005,balitsky.2007,kovchegov.weigert.2007}. We will start by discussing dipole-nucleus scattering, and then, move on to virtual photon-nucleus scattering, which is measurable at colliders. Indeed, the former is more fundamental in the theoretical sense, as the color dipole is a convenient QCD object to represent a virtual photon in DIS at high energy, and it is better controlled theoretically. Furthermore, its characteristics would be manifested in the virtual photon-nucleus scattering since the latter is a mere weighted average of the former over the dipole transverse sizes. Phenomenologically, our purpose is to produce predictions for the distribution of diffractive gaps, which can be extracted from the outputs of future electron-ion colliders, such as BNL-EIC (recently approved)~\cite{EIC.2016} or CERN-LHeC (under design)~\cite{LHeC.2012}.

The paper is organized as follows. In the next section, we will introduce the Kovchegov-Levin formulation of the diffractive DIS and define the quantities of interest for the current analysis. The partonic picture for diffractive dissociation proposed in Refs.~\cite{munier.mueller.2018a,munier.mueller.2018b} from which the rapidity gap distribution can be derived is also briefly reviewed in this section. In~\cref{sec:result}, the numerical results on diffractive gap distributions for the nuclear scattering of both color dipole and virtual photon will be presented. We will then discuss the effects of the running coupling correction to diffraction and compare to available results on the diffractive distributions in~\cref{sec:discussion}. Finally, in~\cref{sec:conclusion}, we will summarize and conclude.
\section{\large FORMALISM}
\label{sec:formalism}

\subsection{Dipole model and Kovchegov-Levin formulation for diffractive dissociation}
\label{subsec:KL_formulation}

At high energy, it is convenient to describe the DIS process in a frame where the virtual photon fluctuates into a $q\bar{q}$ dipole (hereafter referred to as {\em onium}) long before the interaction with the nuclear target. The virtual photon-nucleus interaction is then translated into the interaction between the onium and the nucleus. Diffraction corresponds to the exchange by a color-neutral gluonic state. In this picture, the nuclear scattering cross sections of a virtual photon can be written as a virtuality-dependent weighted average of the scattering cross sections of onia off the nucleus over dipole transverse sizes. The rapidity gap distribution, which is defined as the diffractive cross section at a fixed rapidity gap $Y_0$ normalized to the total inclusive cross section for the scattering of a virtual photon $\gamma^*$ of virtuality $Q^2$ off a nucleus $A$ at a total rapidity $Y=\ln\left[(\hat{s}+Q^2)/Q^2\right]$, where $\hat{s}$ is the squared center-of-mass energy of the scattering process, reads
\begin{equation}
	\mathcal{R}^{\gamma^*A}\equiv\left(-\frac{1}{\sigma_{tot}^{\gamma^*A}}\frac{d\sigma_{diff}^{\gamma^*A}}{dY_0}\right)(Q^2,Y,Y_0) = \frac{\int d^2\underline{r}\int\limits_{0}^{1}  dz \displaystyle\sum_{p=L,T;f}|\psi_{p}^f(\underline{r},z,Q^2)|^2 \ \left[-\frac{d\sigma_{diff}^{q\bar{q}A}}{dY_0}(\underline{r},Y,Y_0)\right]}{\int d^2\underline{r}\int\limits_{0}^{1}  dz \displaystyle\sum_{p=L,T;f}|\psi_{p}^f(\underline{r},z,Q^2)|^2 \ \sigma_{tot}^{q\bar{q}A}(\underline{r},Y)} ,
	\label{eq:photon_gap_distrib}	
\end{equation}
where integrations are performed over all possible transverse sizes $\underline{r}$ of the onium, and over all momentum fractions $z$ of the virtual photon carried by the quark. The probability density functions $|\psi_{L,T}^f(\underline{r},z,Q^2)|^2$ of the quantum fluctuation $\gamma^*\to q_f\bar{q}_f$ in longitudinal (L) and transverse (T) polarizations for a quark flavor $f$ are given by~\cite{nikolaev.zakharov.1991b,bjorken.etal.1971}

\begin{align}
	|\psi_{L}^f(\underline{r},z,Q^2)|^2 & = \frac{\alpha_{\rm em}N_c}{2\pi^2} 4Q^2 z^2(1-z)^2 e_f^2 K_0^2(ra_f), \label{eq:wave_func_L}\\
	|\psi_{T}^f(\underline{r},z,Q^2)|^2 & = \frac{\alpha_{\rm em}N_c}{2\pi^2} e_f^2\left\{a_f^2K_1^2(ra_f)\left[z^2+(1-z)^2\right] + m_f^2K_0^2(ra_f)\right\}, 
	\label{eq:wave_func_T}
\end{align}
where $r = |\underline{r}|$, $a_f^2 = Q^2z(1-z) + m_f^2$, and $m_f$ and $e_f$ are the mass and the charge of a quark of flavor $f$, respectively. The total onium-nucleus cross section $\sigma_{tot}^{q\bar{q}A}$ is related to the forward elastic scattering amplitude $N$ by
\begin{equation}
	\sigma_{tot}^{q\bar{q}A}(\underline{r},Y) = \sigma_02N(r,Y),
	\label{eq:tot_dip_cross_sec}
\end{equation}
where we assume impact parameter independence of $N$ in such a way that the impact-parameter integration results in an overall dimensionful parameter $\sigma_0$. The diffractive cross sections $\sigma_{diff}^{q\bar{q}A}(r,Y,Y_0)$ and $\sigma_{diff}^{\gamma^*A}(Q^2,Y,Y_0)$ for the nuclear scattering of the onium and the virtual photon, respectively, are defined for a minimal rapidity gap $Y_0$. Assuming again the impact parameter independence, $\sigma_{diff}^{q\bar{q}A}$ can be expressed as
\begin{equation}
	\sigma_{diff}^{q\bar{q}A}(\underline{r},Y,Y_0) =  \sigma_0 N_D(r,Y,Y_0),
	\label{eq:diff_dipole_nucl_cs}
\end{equation}
where the function $N_D(r,Y,Y_0)$ represents the diffractive onium-nucleus cross section per unit impact parameter with the minimal gap $Y_0$.

The knowledge on the onium-nucleus scattering profiles $N$ and $N_D$ is then essential to the investigation of diffractive patterns. In the dipole picture, the onium may further evolve by QCD radiations to a complex Fock state at the interaction time. At low Bjorken-$x$ ($Y=\ln(1/x)$) and large number of colors $N_c$, the QCD evolution of the elastic amplitude $N$ of an onium of size $r$ with a nucleus in rapidity $Y$ is governed, at leading order, by the Balitsky-Kovchegov (BK) nonlinear equation~\cite{balitsky.1996,kovchegov.1999}
\begin{equation}
	\frac{\partial N}{\partial Y}(r,Y) = \int d^2\underline{r}_1 K^{LO}(r, r_1,r_2) \left[N(r_1,Y)+N(r_2,Y)-N(r,Y)-N(r_1,Y)N(r_2,Y)\right],
	\label{eq:BK_LO}
\end{equation}
with the leading-order kernel $K^{LO}(r, r_1,r_2)$ given by~\cite{mueller.1994}
\begin{equation}
	K^{LO}(r, r_1,r_2) = \frac{\bar{\alpha}_s}{2\pi}\frac{r^2}{r_1^2 r_2^2},
	\label{eq:LO_kernel}
\end{equation}
where the QCD coupling $\bar{\alpha}_s \equiv \frac{\alpha_sN_c}{\pi}$ is kept fixed. The initial condition for the BK equation is assumed to be given by the McLerran-Venugopalan (MV) amplitude~\cite{mclerran.venugopalan.1994a,mclerran.venugopalan.1994b}:
\begin{equation}
	N_{MV} (r,Y=0) = 1 - \exp\left[-\frac{r^2Q_{A}^2}{4}\ln\left(e + \frac{1}{r^2\Lambda_{QCD}^2}\right)\right],
	\label{eq:MV}
\end{equation}
where $Q_{A}$ is the nuclear saturation momentum at zero rapidity, which encodes the nuclear ($A$) dependence of the amplitude $N$ (see the Appendix). 

Meanwhile, the diffractive cross section $N_D$ was found to obey the leading-order Kovchegov-Levin (KL) equation~\cite{kovchegov.levin.2000}, valid in the same limits to the BK equation (\ref{eq:BK_LO}), which is expressed as
\begin{equation}
	\begin{split}
	\frac{\partial N_D}{\partial Y}(r,Y,Y_0) = \int d^2\underline{r}_1 & K^{LO}(r,r_1,r_2) \left[ N_D(r_1,Y,Y_0) + N_D(r_2,Y,Y_0) - N_D(r,Y,Y_0)\right. \\
	& \left. - 2N(r_1,Y)N_D(r_2,Y,Y_0) - 2N(r_2,Y)N_D(r_1,Y,Y_0) \right.\\
	& \left. + N_D(r_1,Y,Y_0)N_D(r_2,Y,Y_0) + 2N(r_1,Y)N(r_2,Y)\right],
	\end{split}
	\label{eq:KL_LO}
\end{equation}
with the initial condition set at $Y_0$:
\begin{equation}
	N_D(r,Y=Y_0,Y_0) = N^2(r,Y_0).
	\label{eq:ND_init}
\end{equation}
For the sake of convenience for the numerical calculation, we introduce the cross section per impact parameter $N_{in}$ defined by 
\begin{equation}
	N_{in}(r,Y,Y_0) = 2N(r,Y) - N_D(r,Y,Y_0),
	\label{eq:N_in}
\end{equation}
which encodes all inelastic contributions to the scattering. It is straightforward to show that $N_{in}$ also satisfies the leading-order BK equation (\ref{eq:BK_LO}). From \cref{eq:ND_init,eq:N_in}, the initial condition for $N_{in}$ reads 
\begin{equation}
	N_{in}(r,Y=Y_0,Y_0) = 2N(r,Y_0) -  N^2(r,Y_0).
	\label{eq:Nin_init}
\end{equation}
While the BK equation is known at next-to-leading order~\cite{balitsky.2007,kovchegov.weigert.2007,balitsky.chirilli.2008,Ducloue.etal.2019}, the KL equation beyond the leading order has not been established. The only known subleading correction to the KL equation comes from the running of the strong coupling~\cite{kovchegov.2012}, which is known as one of the largest corrections to the color dipole evolution. To include such correction, one replaces the leading-order kernel (\ref{eq:LO_kernel}) by a theoretical-motivated running coupling kernel $K^{rc}$. Different prescriptions have been proposed~\cite{balitsky.2007,kovchegov.weigert.2007,albacete.kovchegov.2007}. In the current analysis, we consider the following ones:
\begin{enumerate}[label=(\roman*)]
	\item the Balitsky prescription~\cite{balitsky.2007} in which the kernel $K^{LO}$ is replaced by the following expression:
\begin{equation}
	K^{rc} \equiv K^{Bal}(r, r_1,r_2) = \frac{\bar{\alpha}_s(r^2)}{2\pi}\left[ \frac{r^2}{r_1^2r_2^2} + \frac{1}{r_1^2} \left( \frac{\bar{\alpha}_s(r_1^2)}{\bar{\alpha}_s(r_2^2)} - 1\right) + \frac{1}{r_2^2} \left( \frac{\bar{\alpha}_s(r_2^2)}{\bar{\alpha}_s(r_1^2)} - 1\right) \right],
	\label{eq:Balitsky_kernel}
\end{equation}
	\item the so-called ``parent dipole" prescription~\cite{albacete.kovchegov.2007} in which the coupling runs with the size $r$:
\begin{equation}
	K^{rc} \equiv K^{pd} (r, r_1,r_2) = \frac{\bar{\alpha}_s(r^2)}{2\pi}\frac{r^2}{r_1^2r_2^2}.
	\label{eq:pd_kernel}
\end{equation}
\end{enumerate}
We follow Refs.~\cite{albacete.kovchegov.2007,albacete.etal.2009} to regularize the running coupling $\bar{\alpha}_s(r^2)$ to avoid the Landau pole. In particular, for dipole sizes under some threshold $r \le r_{thres}$, the coupling is computed from the following expression:
\begin{equation}
	\bar{\alpha}_s(r^2) = \frac{12N_c}{(11N_c-2N_f)\ln\left(\frac{4C^2}{r^2\Lambda_{\rm QCD}^2}\right)},
	\label{eq:rc_alpha}	
\end{equation} 
where the number of quark flavors $N_f$ and the number of colors $N_c$ are fixed at values $N_f=3$ and $N_c=3$, respectively. The constant $C$ reflects the uncertainty in the Fourier transform from momentum space to coordinate space. Meanwhile, for larger dipole sizes, $r > r_{thres}$, the coupling is frozen to a fixed value $\bar{\alpha}_{thres}$ defined by $\bar{\alpha}_{thres} \equiv \bar{\alpha}_s(r_{thres}^2)$. 

In terms of $N_D$ or of $N_{in}$, and of $N$, the distribution of rapidity gaps in diffractive onium-nucleus scattering can be written as
\begin{equation}
	\mathcal{R}^{dip} \equiv -\frac{1}{2N}\frac{\partial N_D}{\partial Y_0} = \frac{1}{2N}\frac{\partial N_{in}}{\partial Y_0}.
	\label{eq:dip_gap_distrib} 
\end{equation}

\subsection{Diffractive dissociation of small onia off nuclei from a partonic model}
\label{subsec:partonic_model}
We are now going to review the recently proposed partonic model for diffractive dissociation, which is mainly based on Refs.~\cite{munier.mueller.2018a,munier.mueller.2018b,mueller.munier.2014}.

We consider the nuclear scattering of an onium of size $r$ at a large total rapidity Y ($Y\gg 1$), in a frame such that the original onium is boosted to a rapidity $\tilde{Y}$ ($0<\tilde{Y}\le Y$). In such frame, the BK evolution admits a statistical interpretation. Let us start by the realization of the high energy evolution on the wave function of the onium. In one step of evolution, a soft gluon can appear in the onium wave function, as being emitted from either the quark or the antiquark. This single soft gluon emission is, at large $N_c$, tantamount to a branching from one parent dipole ($r$) to two daughter dipoles ($r_1$ and $r_2$) whose probability, at leading order, is given by \cref{eq:LO_kernel}~\cite{mueller.1994}. Such dipole branching is then iterated in the course of evolution. As a result, the Fock state of the onium at the rapidity $\tilde{Y}$ appears as a stochastic set of dipoles with various transverse sizes. 

We now define $P(r,\tilde{Y}|R)$ as the probability of having at least one dipole larger than some size $R$ in the wave function of the initial onium of size $r$ evolved to the rapidity $\tilde{Y}$~\cite{mueller.munier.2014}. It is straightforward to show that the BK equation (\ref{eq:BK_LO}) also controls the rapidity evolution of the probability $P(r,\tilde{Y}|R)$, with the intial condition given by $P(r,\tilde{Y}=0|R) = \theta\left[\ln (r^2/R^2)\right]$. If we further assume the onium is small such that
\begin{equation}
	1 \ll \ln\frac{2}{rQ_s(Y)} \ll \sqrt{Y}, 
	\label{eq:scaling_windows}
\end{equation}
then $P(r,\tilde{Y}|X)$ and $N(r,Y)$ have the same functional form at asymptotic high rapidities, up to appropriate substitutions~\cite{mueller.munier.2014}. In \cref{eq:scaling_windows}, $Q_s(Y)$ is the nuclear saturation scale at rapidity $Y$. The region defined by \cref{eq:scaling_windows} is called as the ``scaling region", since in this region, the total cross section is effectively a function of the scaling variable $\ln\left[2/\left(rQ_s(Y)\right)\right]$ only, which property is referred to as {\em geometric scaling} \cite{Stasto.etal.2001}. 

Let us now consider a particular frame in which the nucleus is boosted to a rapidity $Y_0 > 0$, and the onium evolves to the remaining rapidity $\tilde{Y_0}\equiv Y-Y_0$. Taking the initial onium to be smaller than the inverse saturation scale $2/Q_s(Y)$, to have a diffractive event with a significant probability, there should be a fluctuation which creates at least one large dipole whose size is larger than the inverse saturation scale $2/Q_s(Y_0)$ in the onium wave function at $\tilde{Y_0}$. Such dipole will scatter off the nucleus with a probability of order unity, and with a high fraction ($\sim 1/2$) for elastic processes in which the nucleus is kept intact. In addition, since different dipoles in the onium's Fock state at $\tilde{Y_0}$ interact differently with the nucleus, the onium will be dissociated into particles in the final state. We then have a diffractive event with rapidity gap $Y_0$ measured from the nucleus.

In the spirit of that picture, the diffractive cross section with fixed gap $Y_0$ is proportional to the probability $P(r,\tilde{Y_0}|2/Q_s(Y_0))$. The rapidity gap distribution for the diffractive dissociation of onia of size in the scaling window (\ref{eq:scaling_windows}) is then expressed as~\cite{munier.mueller.2018a,munier.mueller.2018b}
\begin{equation}
	\mathcal{R}^{dip}_{asymp} = c_{D} \left[\frac{Y}{Y_0(Y-Y_0)}\right]^{3/2},
	\label{eq:theoretical_gap_distrib}
\end{equation}
which is valid for $Y,Y_0,Y-Y_0 \gg 1$. The constant $c_D$ is undetermined from the above-mentioned model. However, we have recently shown this constant reads ~\cite{in.prep}
\begin{equation}
	c_D = \frac{1}{\sqrt{\bar{\alpha}_s}}\frac{\ln 2}{\gamma_0\sqrt{2\pi\chi''(\gamma_0)}}
	\label{eq:constant}
\end{equation}
where $\chi(\gamma) = 2\psi(1)-\psi(1-\gamma)-\psi(\gamma)$, and $\gamma_0$ solves the equation $\chi'(\gamma_0)=\chi(\gamma_0)/\gamma_0$. The distribution (\ref{eq:theoretical_gap_distrib}) indicates diffractive events with small rapidity gaps ($Y_0 \ll Y/2$) or large rapidity gaps ($Y_0 \gg Y/2$) are more probable to occur than events with moderate gaps ($Y_0 \sim Y/2$).

While an asymptotic analytical expression of the rapidity gap distribution is available in the fixed coupling case, there are still no analytical calculations of such quantity when taking into account the running coupling correction. One motivation of the current numerical analysis is to check whether the above-mentioned prediction for the asymptotic behavior of the rapidity gap distribution already manifests at finite rapidities, and whether the running coupling effects could make differences to the predicted diffractive gap pattern in \cref{eq:theoretical_gap_distrib}.

\section{\large NUMERICAL EVALUATION OF DIFFRACTIVE CROSS SECTIONS}
\label{sec:result}
Let us first present our choices of kinematic variables. We select two following values of rapidity: 
\begin{enumerate}[label=(\roman*)]
	\item $Y=6$, or the Bjorken $x$ variable $x=e^{-Y} \approx 0.002$. This value of $Y$ is accessible at BNL-EIC for low to moderate center-of-mass energies~\cite{EIC.2016}, such as $\sqrt{s}=90\ \rm GeV$ or $\sqrt{s}=45\ \rm GeV$ ($A\ge 56$), and at CERN-LHeC for $\sqrt{s_{ePb}}=877\ \rm GeV$ ($70\ \rm GeV$ -- $2.75\ \rm TeV$)~\cite{LHeC.2012}. 
	\item $Y=10$, or $x\approx 4.5\times 10^{-5}$. It is accessible at CERN-LHeC at $\sqrt{s_{ePb}}=877\ \rm GeV$. 
\end{enumerate}

For the photon virtuality, we select pertubative values in the range $Q^2 = 1-10 \rm\ GeV^2$.

We shall start with a detailed study of the diffractive onium-nucleus scattering for different onium sizes since this is the process whose asymptotics was analysed theoretically and since the virtual photon-nucleus cross sections are just onium-nucleus cross sections average over the size of the onium weighted by squared wave function of the virtual photon (see \cref{eq:photon_gap_distrib}). In order to see the convergence toward the asymptotic solution known analytically, we additionally present the results for an unrealistic rapidity $Y=30$. In the second part, we shall address actual observables at an electron-ion collider.

\subsection{Onium-nucleus scattering}

It is convenient to introduce the scaling variable
\begin{equation}
	\tau \equiv \ln\frac{2}{rQ_s(Y)},
	\label{eq:log_dp_size}
\end{equation}
where the saturation momentum $Q_s(Y)$ is defined by the condition $N(r=2/Q_s(Y),Y) = 0.5$. Positive values of $\tau$ parameterize the kinematic region in which the forward elastic scattering amplitude is small ($N \to 0$ when $\tau \to \infty$) that we shall call the ``dilute" region hereafter, while negative values of $\tau$ parameterize the saturation region ($N\to 1$ when $\tau \to -\infty$).

\begin{figure}[h!]
	\centering
  	\includegraphics[width=\linewidth]{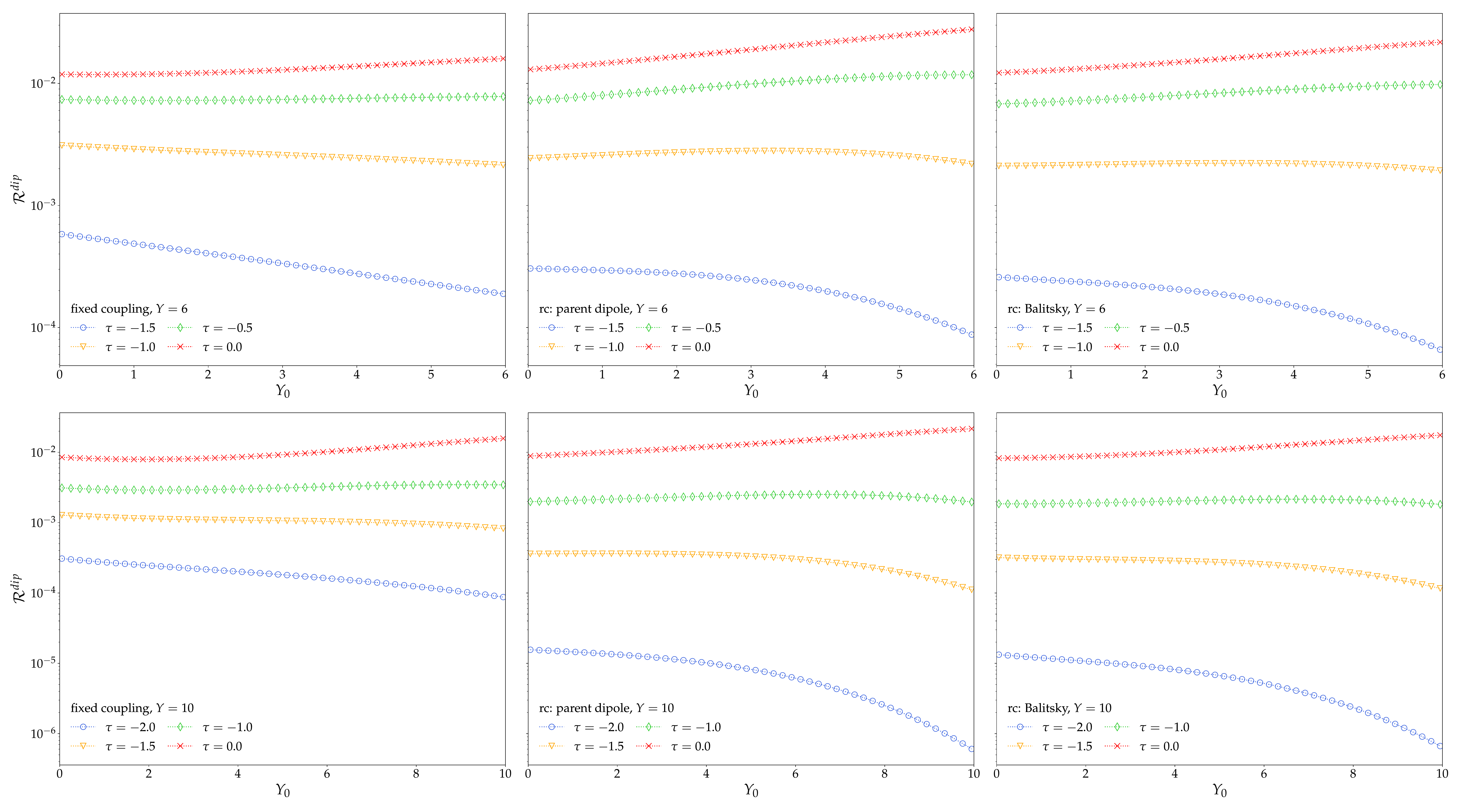}%
  	\caption{Rapidity gap distributions for different onium sizes picked in the saturation region $(\tau \le 0)$ at $Y=6$ (first row) and $Y=10$ (second row) considering three schemes: fixed coupling and two running coupling prescriptions considered in the current analysis.}
	\label{fig:dp_gap_sat}
\end{figure}  

\cref{fig:dp_gap_sat} shows the rapidity gap distributions $\mathcal{R}^{dip}$ for the onium picked in the saturation regime $(\tau \le 0)$. As the onium becomes larger in size, we are closer to the black-disk limit at which there should be an equal probability of $1/2$ of having elastic and inelastic scatterings. These contributions are excluded from the definition of the distribution $\mathcal{R}^{dip}$. Consequently, the contribution from diffractive scattering $\mathcal{R}^{dip}$ is suppressed as $\tau$ becomes more negative. Such suppression is stronger if the running of the strong coupling is included. Additionally, for the onium sizes far enough from the inverse saturation scale, large rapidity gaps are suppressed compared to the small and medium ones. The $\tau \le 0$ regime has not been studied theoretically. 

\begin{figure}[ht!]
	\centering
	\includegraphics[width=\linewidth]{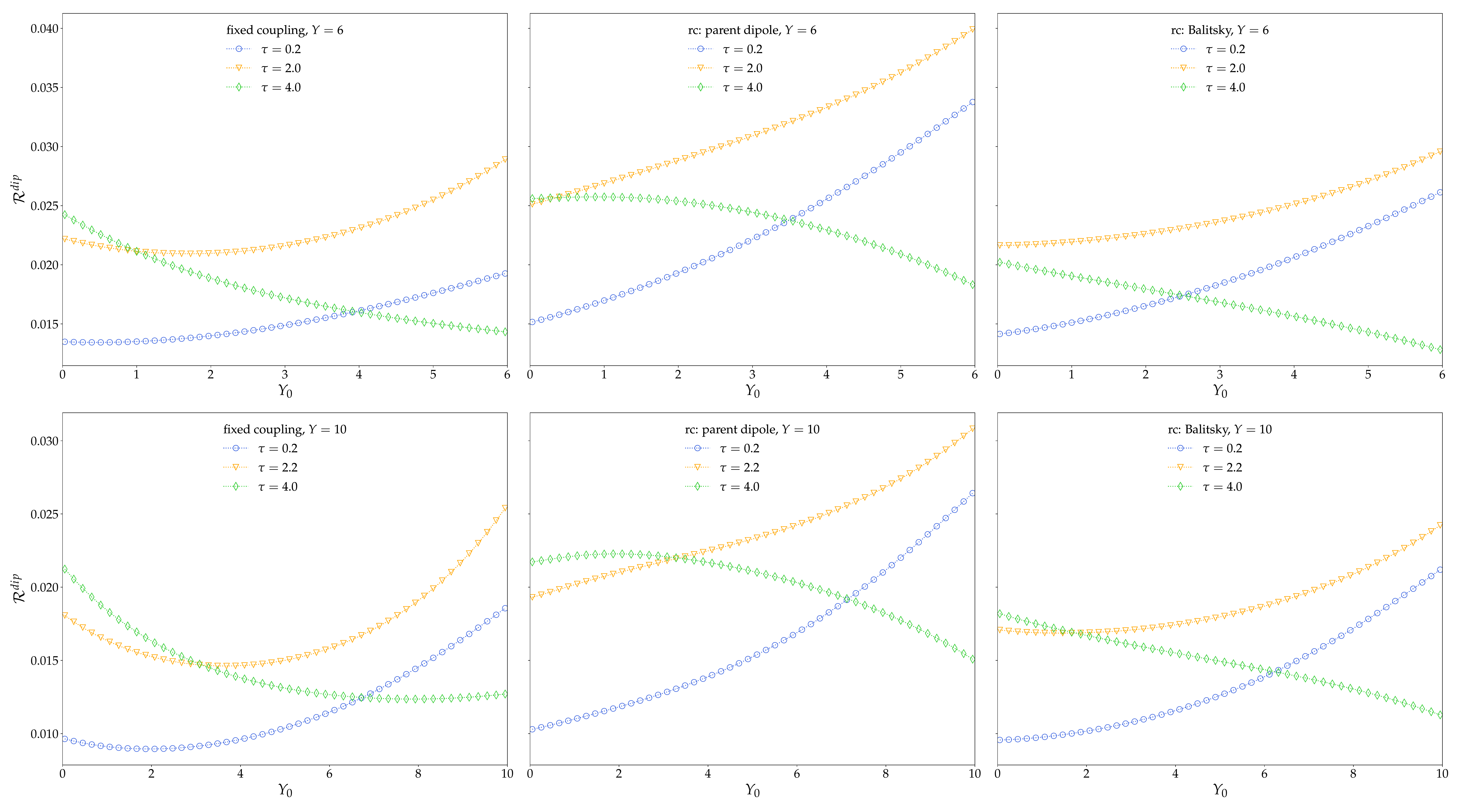}%
	\caption{Rapidity gap distributions for different onium sizes picked in the dilute region ($\tau>0$) for $Y=6$ (first row) and $Y=10$ (second row) considering three schemes: fixed coupling and two running coupling prescriptions considered in the current analysis.}
	\label{fig:dp_gap_dilute}
\end{figure} 

The rapidity gap distributions in the dilute regime ($\tau>0$;~\cref{fig:dp_gap_dilute}) reflect general features of the aforementioned partonic picture for diffractive dissociation, which can be extended to the case of running coupling evolution. For the onium sizes close to the inverse saturation momenta, large-gap diffraction dominates over small-gap diffraction since it is readily to have fluctuations generating dipoles whose size is of order inverse saturation scale at $Y_0$ in the early stages of the evolution of the onium. As the onium becomes smaller, large-dipole splittings are more probably to occur in later stages of the evolution, since the onium needs more rapidity to develop. Consequently, the contribution of small-gap diffraction becomes more important compared to large-gap one. When the onium size approaches the color transparency limit~\cite{nikolaev.zakharov.1991b}, where $N\sim (rQ_A)^2\to 0$ as $r\to 0$, large-dipole fluctuations are less probable, and the contribution of diffractive dissociation to the onium-nucleus scattering is suppressed. At such limit, the contribution from the inelastic scattering would dominate the interaction. 

\begin{figure}[ht!]
	\centering
  	\includegraphics[width=\linewidth]{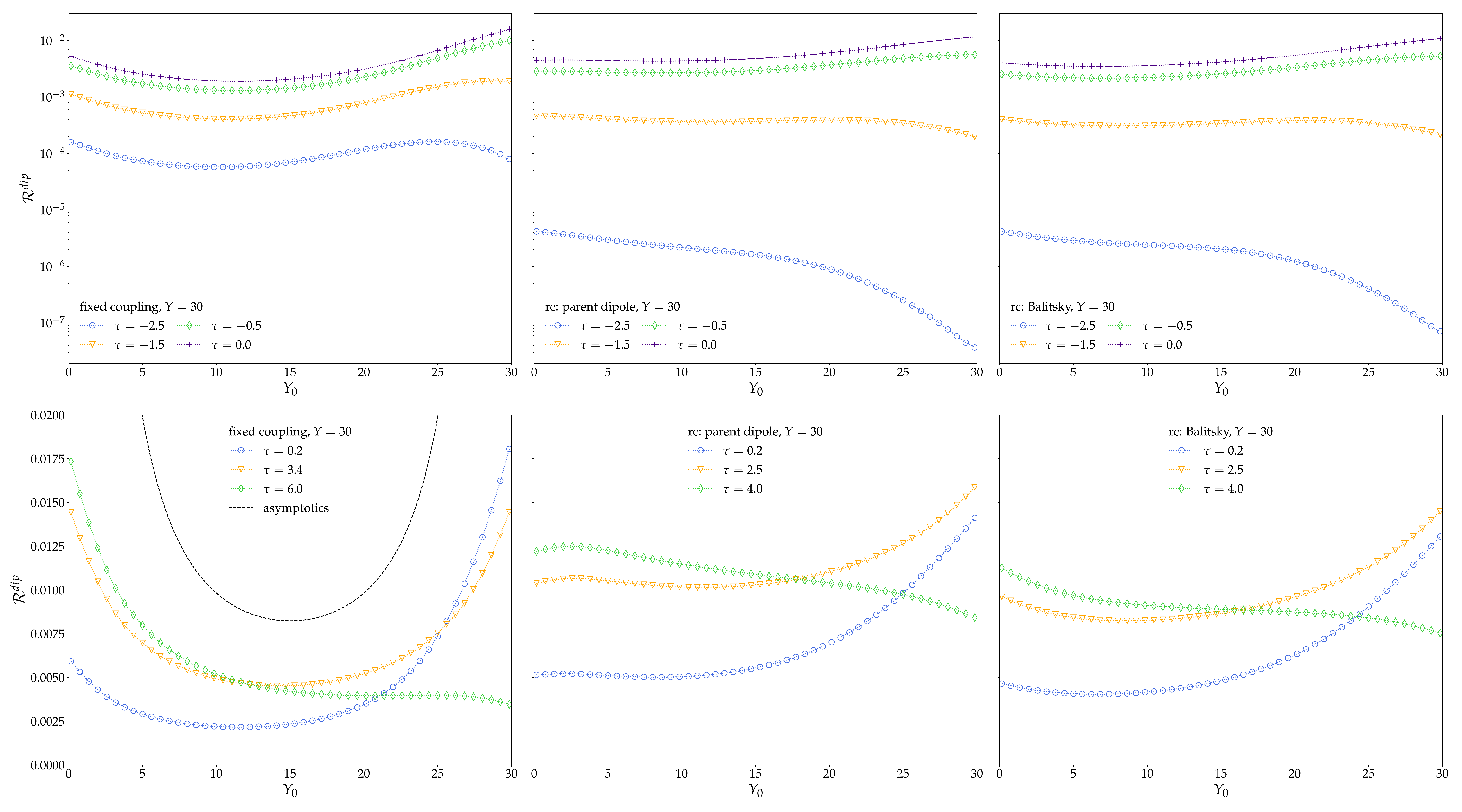}  
	\caption{Rapidity gap distributions for different onium sizes picked in the saturation region (first row) and in the dilute region (second row) for the total rapidity $Y=30$ considering three schemes: fixed coupling and two running coupling prescriptions considered in the current analysis. In the case of fixed coupling and $\tau>0$, the asymptotic prediction in \cref{eq:theoretical_gap_distrib} with the constant $c_D$ in \cref{eq:constant} is superimposed (the black dashed curve).}
	\label{fig:dsigma_dp_30}
\end{figure}

Considering the fixed coupling case, the hammock shape predicted by the partonic picture of diffraction is already exhibited at realistic values of rapidity ($Y=6$ and $Y=10$; see~\cref{fig:dp_gap_dilute}). In order to check that this peculiar shape corresponds indeed to the onset of the asymptotics in~\cref{eq:theoretical_gap_distrib}, we push the calculation to a higher value of the total rapidity $Y$, in particular $Y=30$ (see~\cref{fig:dsigma_dp_30}), even though this rapidity is not accessible at planned electron-ion colliders. We see that the distributions in the scaling region have a similar shape to the predicted asymptotics (see the plot for $\tau = 3.4$). However, finite-rapidity corrections are significant even at such rapidity, which screen the asymptotic behavior. Furthermore, we see that the hammock shape also presents in the distributions with the running coupling correction, or for the onium sizes in the saturation region and close to the inverse saturation scale. 

In summary, the rapidity gap distribution for the diffractive onium-nucleus scattering depends on the regime (either saturation or dilute) where the onium is picked, and on its relative size compared to the inverse saturation scale. The running of the QCD coupling modifies the distribution in comparison to the fixed coupling case, however it is not susceptible to the selection of the running coupling prescription. In the case of fixed coupling, the rapidity gap distributions for the onium sizes in the dilute regime and not very far from the inverse saturation scale exhibit the peculiar shape predicted by the aforementioned partonic picture for diffractive dissociation. 

\subsection{Virtual photon-nucleus scattering}

\begin{figure}[ht!]
	\centering
  	\includegraphics[width=\linewidth]{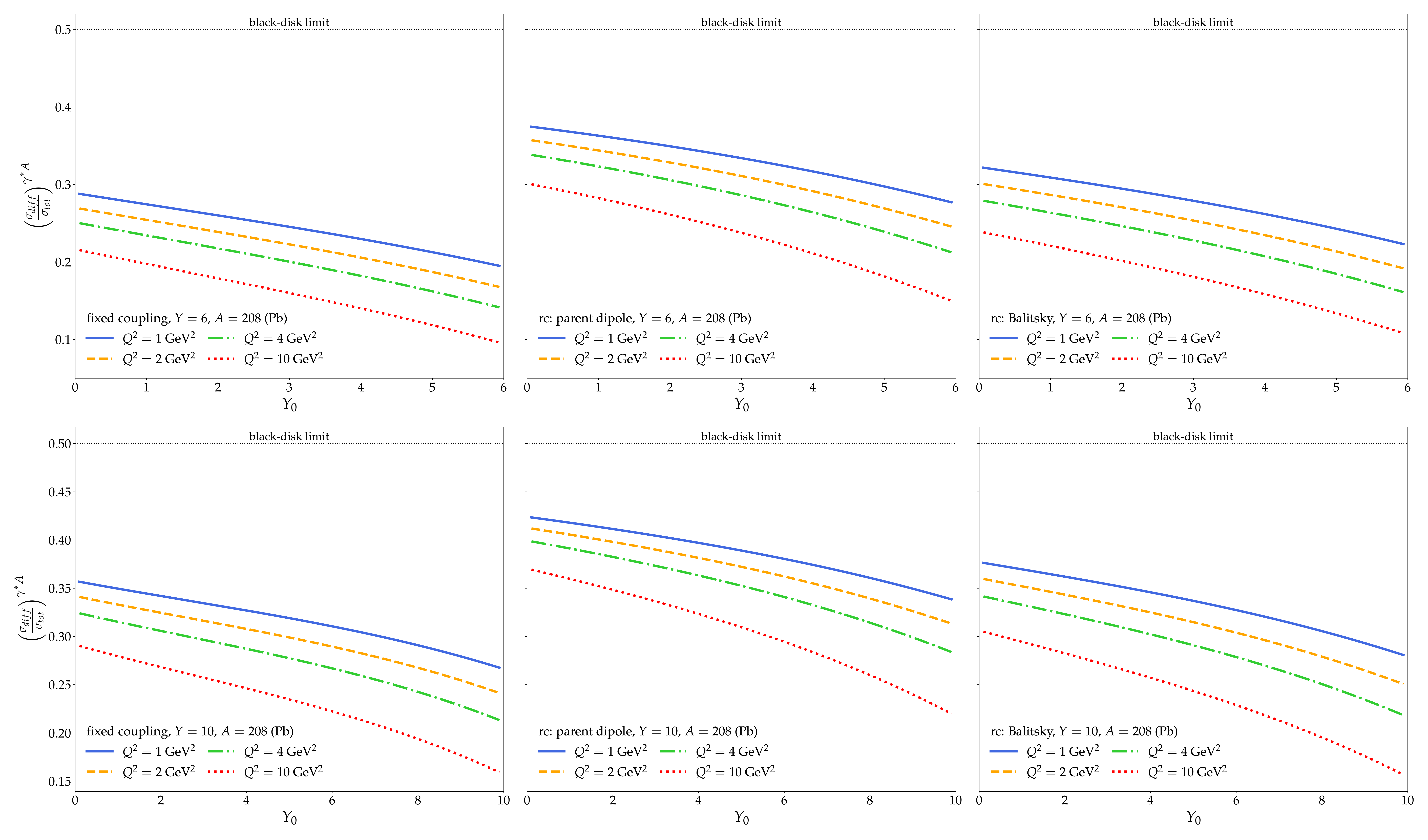}  
	\caption{Dependence of the ratio $\left(\sigma_{diff}/\sigma_{tot}\right)^{\gamma^*A}$ on the minimal rapidity gap $Y_0$ and on the virtuality $Q^2$ at two values of the total rapidity $Y=6$ (first row) and $Y=10$ (second row) considering three schemes: fixed coupling and two running coupling prescriptions considered in the current analysis.}
	\label{fig:sigma_diff_tot}
\end{figure}

Let us start with the diffractive cross section with a {\em minimal rapidity gap $Y_0$} normalised to the total cross section:
\begin{equation}
	\left(\frac{\sigma_{diff}}{\sigma_{tot}}\right)^{\gamma^*A}(Q^2,Y,Y_0) = \frac{\int d^2\underline{r}\int\limits_{0}^{1}  dz \displaystyle\sum_{p=L,T;f}|\psi_{p}^f(\underline{r},z,Q^2)|^2 \ \sigma_{diff}^{q\bar{q}A}(\underline{r},Y,Y_0)}{\int d^2\underline{r}\int\limits_{0}^{1}  dz \displaystyle\sum_{p=L,T;f}|\psi_{p}^f(\underline{r},z,Q^2)|^2 \ \sigma_{tot}^{q\bar{q}A}(\underline{r},Y)},
\end{equation}
which estimates how close to the black-disk limit we are. As shown in \cref{fig:sigma_diff_tot}, the ratio varies slowly with respect to the virtuality $Q^2$ and is closer to its black-disk limit value $0.5$ at lower $Q^2$, since larger onium sizes close to saturation are more probable to be probed. It also depends on the total rapidity $Y$, which becomes larger at a higher total rapidity $Y$. For example, for $Q^2 = 4 \rm\ GeV^2$ and fixed coupling, diffractive events are predicted to account for about $15-25\%$ at $Y=6$ and about $23-32\%$ at $Y=10$ of total scattering events, depending on the value of the minimal gap used as the threshold to probe diffractive events in practice. Such contribution is estimated to be higher by a few percent when taking into account the running of the strong coupling.

\begin{figure}[h!]
	\centering
	\includegraphics[width=\linewidth]{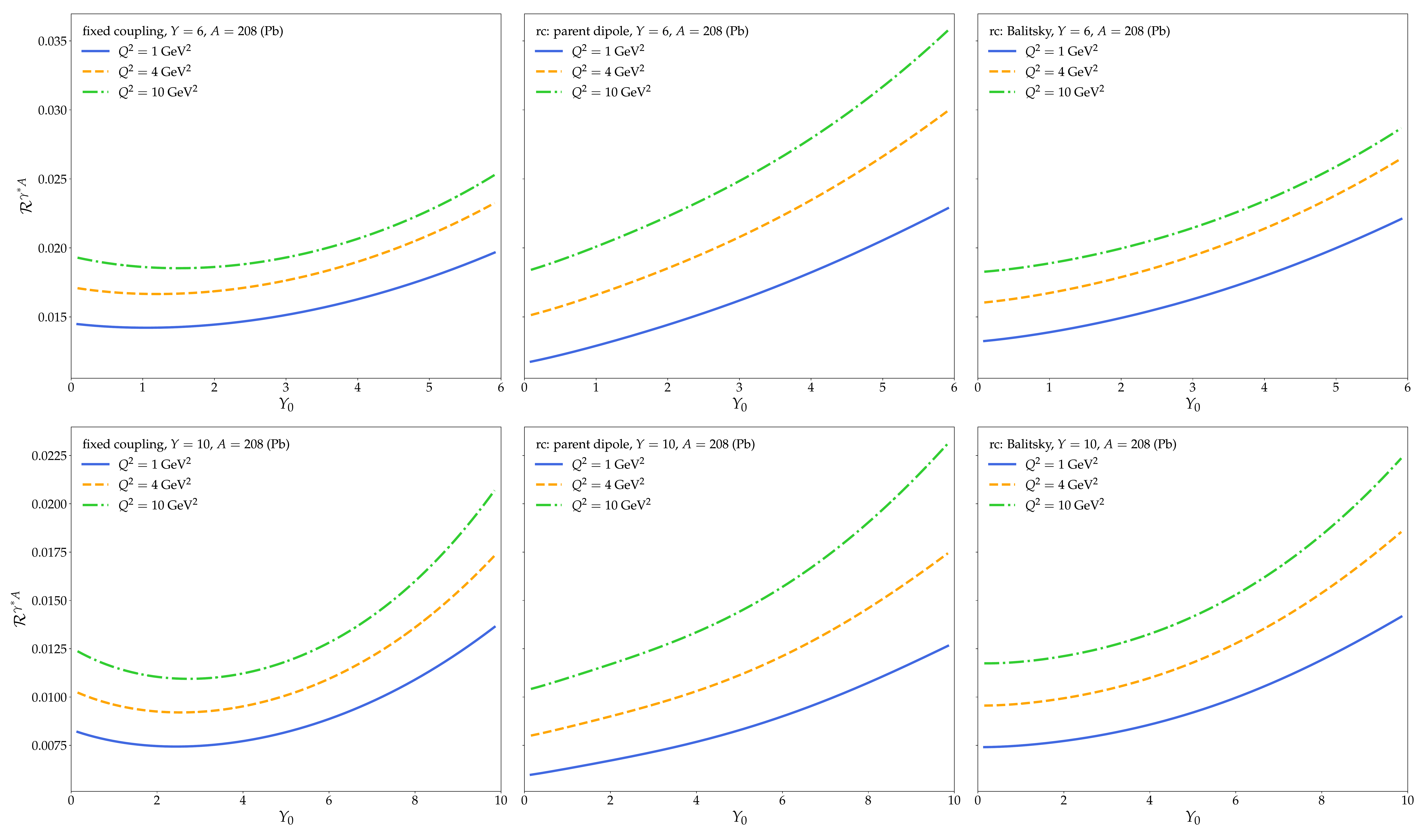}  
	\caption{Dependence of the rapidity gap distributions on the virtuality $Q^2$ at two values of the total rapidity $Y=6$ (first row) and $Y=10$ (second row) considering three schemes: fixed coupling and two running coupling prescriptions considered in the current analysis.}
	\label{fig:dsigma_photon_distrib}
\end{figure}

\begin{figure}[h!]
	\centering
	\includegraphics[width=\linewidth]{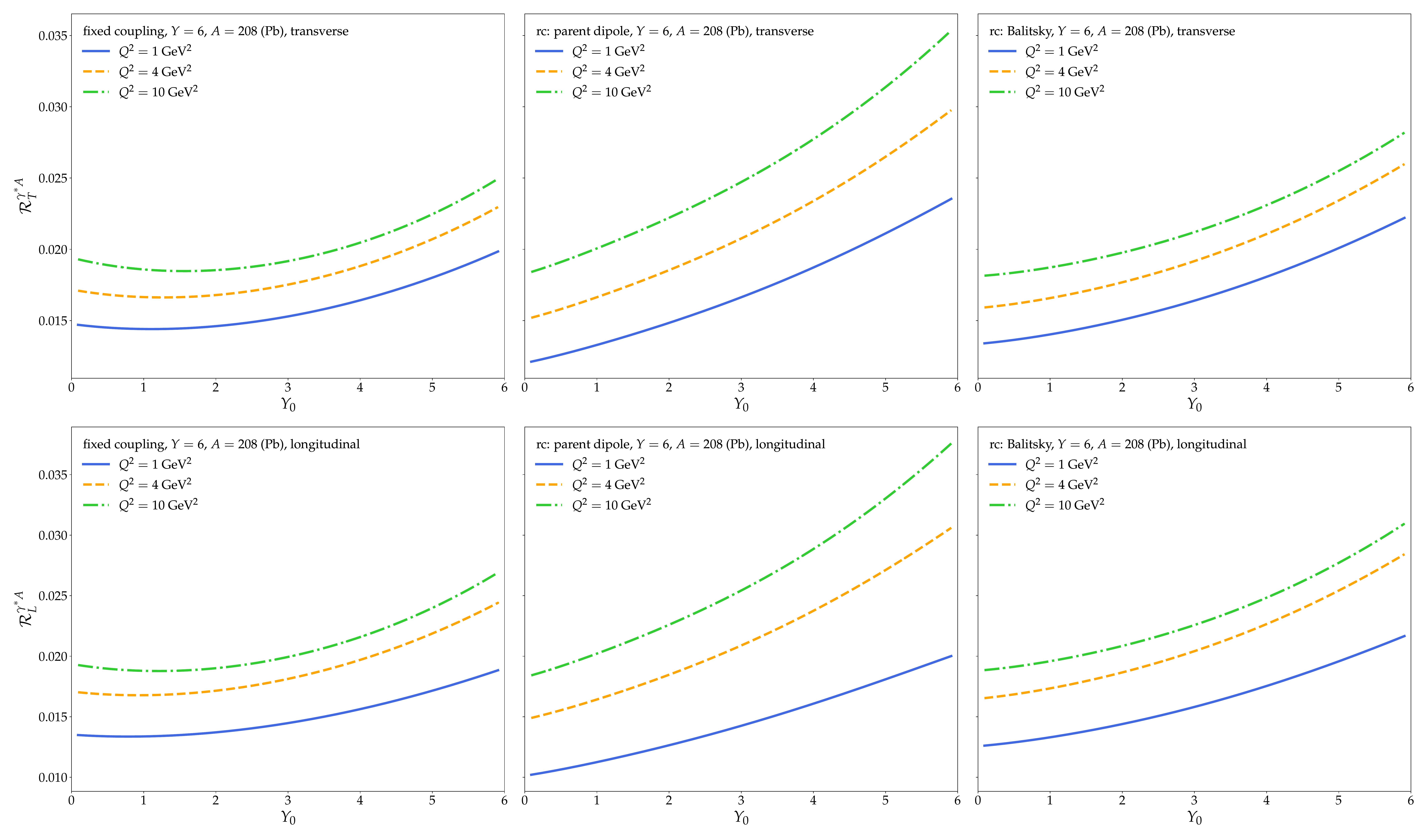}  
	\caption{Rapidity gap distributions considering only the tranverse (first row) or the longitudinal (second row) polarization at the total rapidity $Y=6$ considering three schemes: fixed coupling and two running coupling prescriptions considered in the current analysis.}
	\label{fig:dsigma_photon_LT_6}
\end{figure}

\cref{fig:dsigma_photon_distrib} presents the dependence of the rapidity gap distributions on the virtuality $Q^2$. With the chosen set of the virtuality $Q^2$, the quantity $\ln(Q/Q_s)$, which is the typical value of the pre-defined scaling variable $\tau$, varies in the range $-0.35\lesssim \ln(Q/Q_s)\lesssim 1.34$ for $Y=6$ and $-1.1\lesssim \ln(Q/Q_s)\lesssim 0.82$ for Y=10, considering both fixed and running coupling cases. Therefore, the dominant contribution should come from onium sizes in the vicinity of the inverse saturation scales, as shown by the similarity between the shapes of the obtained distributions and of the distributions in such regime presented previously. As $Q^2$ increases, smaller onia are more probable to be probed, and the scattering gets away from the black-disk limit. Consequently, the distributions become larger and acquire a larger contribution from the dilute regime. Notice that, for large $Q^2$ such that $Q \gg Q_s$, the dominant contribution comes from small onia approaching the color transparency limit, and hence the distribution should be suppressed. Such behaviors are also seen in the rapidity gap distributions considering only the transverse or longitudinal polarization of the virtual photon shown in ~\cref{fig:dsigma_photon_LT_6}. We see that two polarizations give comparable contributions.

\begin{figure}[h!]
	\centering
	\includegraphics[width=\linewidth]{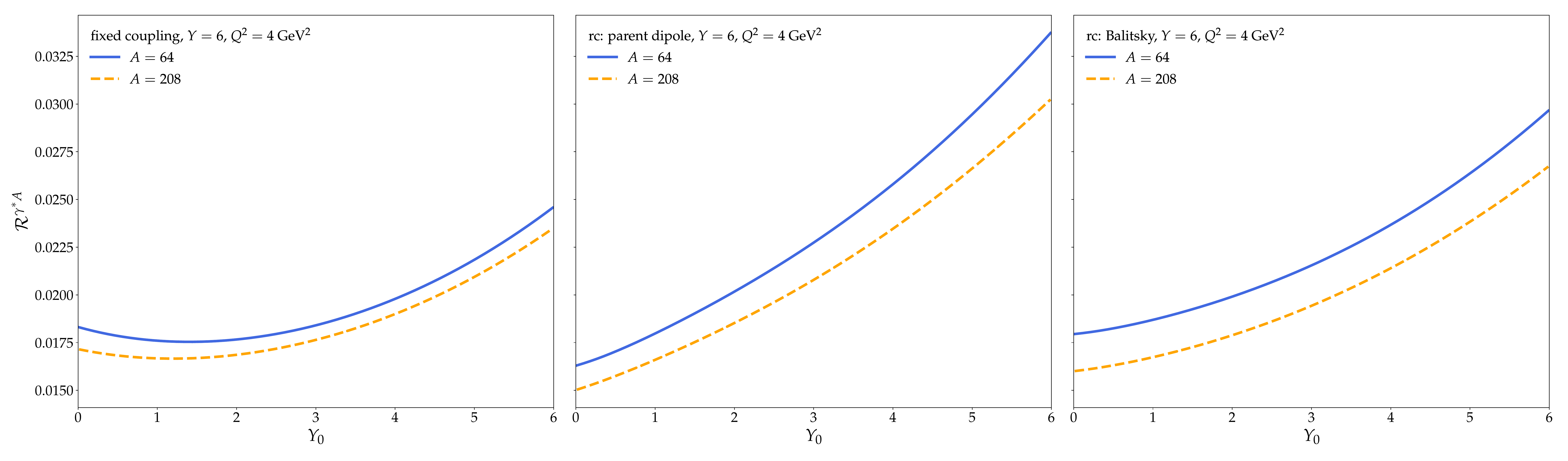}  
	\caption{Rapidity gap distributions at the total rapidity $Y=6$ and the virtuality $Q^2=4 \rm\ GeV^2$ for two different nuclei $A=208$ and $A=64$ considering three schemes: fixed coupling and two running coupling prescriptions considered in the current analysis.}
	\label{fig:diffraction_6_A}
\end{figure}

In \cref{fig:diffraction_6_A}, the rapidity gap distributions for two large nuclei with a particular choice of kinematics are plotted. With a larger nucleus, the initial nuclear saturation momentum, and hence saturation momenta at higher rapidities, become larger. Therefore, the typical scaling variable $\ln(Q/Q_s)$ becomes smaller, and the distributions are suppressed. This suppression is weak, due to the fact that the initial saturation momentum is a slowly varying function of $A$, $Q_A^2 \sim A^{1/3}$ (see the Appendix).

\begin{figure}[h!]
	\centering
	\includegraphics[width=\linewidth]{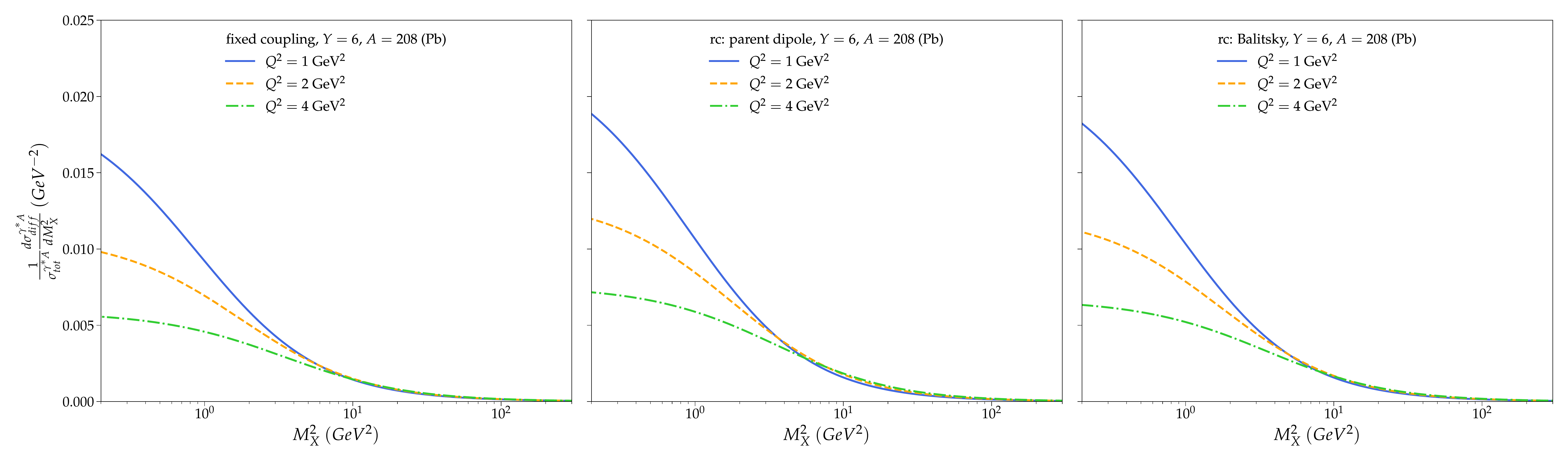}  
	\caption{Dependence of the diffractive mass spectra on the virtuality $Q^2$ at the total rapidity $Y=6$.}
	\label{fig:diff_MX_6}
\end{figure}

We can translate the distribution of the rapidity gap $Y_0$ into the distribution of the (squared) invariant mass $M_X^2$ of the diffractive inclusive final state $X$, which is referred to as diffractive mass spectrum. For this purpose, we employ the relation $Y_0 = Y-\ln\frac{M_X^2+Q^2}{Q^2}$. The diffractive mass spectrum is then related to the rapidity gap distribution as follows:
\begin{equation}
	\frac{1}{\sigma^{\gamma^*A}_{tot}}\frac{d\sigma^{\gamma^*A}_{diff}}{dM_X^2} = \frac{\mathcal{R}^{\gamma^*A}}{M_X^2+Q^2}.
	\label{eq:MX2}
\end{equation}
The mass spectra at the total rapidity $Y=6$ are shown in~\cref{fig:diff_MX_6} for diffrent values of the photon virtuality $Q^2$.  We see that the low-mass regime dominates over the high-mass one, which basically agrees with the general behavior of the obtained rapidity gap distributions that large-gap diffraction is preferred. The above-mentioned relation between the rapidity gap $Y_0$ and the diffractive mass $M_X^2$ has a peculiar consequence: while there is a minimum in the rapidity gap distributions, no local minimum is manifested in the mass spectra. Furthermore, when $Q^2$ increases, the interaction becomes more inelastic as smaller onia are more likely to be probed. Consequently, the low-mass regime is suppressed, and it is more possible for the virtual photon to be dissociated into a high-mass system in the final state at higher $Q^2$. 

\section{\large DISCUSSION}
\label{sec:discussion}

\subsection{Effects of the running coupling correction to diffractive dissociation}

A generic effect of the running of the strong coupling is to suppress the emission of color dipoles with small transverse size in the wave function of the onium, which slows down the small-$x$ evolution. As a consequence, one would expect to have smaller dipoles in the typical Fock state evolved from an initial onium at the time of interaction  when the coupling is fixed than when the running of the coupling is included. Therefore, in the running coupling case, there should be a larger portion of dipoles in the wave function of an onium such that $\tau<0$ located in the saturation region. The scattering is then more elastic and we are closer to the black-disk limit, which results in a stronger suppression of the rapidity gap distributions $\mathcal{R}^{dip}$ at $\tau<0$. 

The running of the coupling also causes the evolution of a small onium to be generally slower than a large one, as it reduces the branching rate when the size of the parent dipole becomes smaller. A consequence of such effect is to diminish the contribution of the small-onium component (or the dilute domain) to the diffractive dissociation of the virtual photon. Therefore, saturation effects become more significant, and the ratio $\left(\sigma_{diff}/\sigma_{tot}\right)^{\gamma^*A}$ is closer to its black-disk limit value when the running of the strong coupling is taken into account, even if the running coupling saturation momentum is smaller than the fixed coupling one (see~\cref{fig:sigma_diff_tot}).

We can extend heuristically the partonic picture for diffractive dissociation presented in \cref{subsec:partonic_model} when the coupling is fixed to the running coupling case at a large value of the total rapidity, $Y \gg 1$. However one should first notice the following essential difference between two cases. If the coupling is fixed, the nuclear saturation scale at a rapidity $Y_0$ is given by~\cite{mueller.trianta.2002, munier.peschanski.2004a}
\begin{equation}
	Q_{s,fc}^2(Y_0) = Q_A^2\exp\left[\bar{\alpha}_s\chi'(\gamma_0) Y_0-\frac{3}{2\gamma_0}\ln(\bar{\alpha}_sY_0)\right],
	\label{eq:qs_fc}
\end{equation}
and the size of the largest dipole in the typical evolution configration of an onium of size $r$ at the rapidity $\tilde{Y}_0 \equiv Y-Y_0$ is~\cite{munier.mueller.2018a}
\begin{equation}
	R_{s,fc}^2(\tilde{Y}_0) = r^2\exp\left[\bar{\alpha}_s\chi'(\gamma_0) \tilde{Y}_0-\frac{3}{2\gamma_0}\ln(\bar{\alpha}_s\tilde{Y}_0)\right].
	\label{eq:rs_fc}
\end{equation}
The similarity between $Q_{s,fc}$ and $R_{s,fc}$ is due to the twofold interpretation of the BK equations: it describes the evolution not only of the elastic scattering amplitude but also of a statistical measure for the region of large dipole sizes (see \cref{subsec:partonic_model}). We can see that, when the onium is picked in the dilute domain ($\tau>0$) and $Y_0, \tilde{Y}_0 \gg 1$, then $\left[Q_{s,fc}^2(Y_0)R_{s,fc}^2(\tilde{Y}_0)\right]/4 < 1$, namely all dipoles constituting the typical evolution configuration of the onium at the rapidity $\tilde{Y}_0$ are smaller than the inverse saturation scale $2/Q_{s,fc}(Y_0)$. In addition, the $Y_0$ dependences of the leading terms in the exponents in \cref{eq:qs_fc,eq:rs_fc} cancel each other, so the rapidity gap distribution is predominantly shaped by the logarithmic terms in the exponents. 

\begin{figure}[h!]
	\centering
	\includegraphics[width=\linewidth]{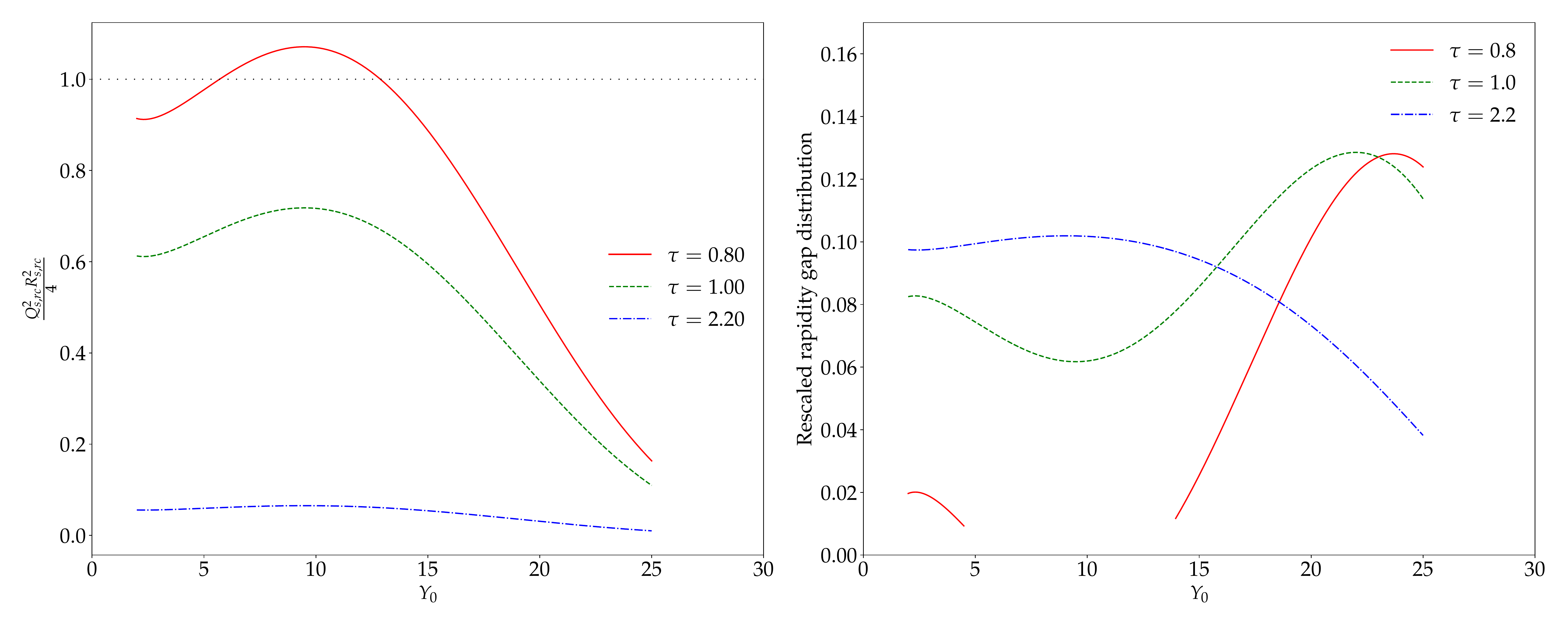}  
	\caption{(Left) Value of the quantity $\left[Q_{s,rc}^2(Y_0)R_{s,rc}^2(\tilde{Y}_0)\right]/4$ as a function of the rapidity gap $Y_0$ at different onium sizes in the running coupling case. This is larger than $1$ if the typical configuration of the onium at $\tilde{Y}_0$ overlaps with the saturation region of the nucleus evolved to $Y_0$. (Right) The rapidity gap distribution rescaled by the total cross section $(2N)$ (the overall constant is set to $1$) at different onium's sizes. Here we require that~\cref{eq:Rdip_rc_theory} is applicable if $\left[Q_{s,rc}^2(Y_0)R_{s,rc}^2(\tilde{Y}_0)\right]/4<0.96$. The total rapidity is $Y=30$, and the correction terms $\delta_{1,2}$ are set to values $6.5$ and $1.1$, respectively.}
	\label{fig:theory}
\end{figure}

The situation is different in the running coupling case. Indeed, the running of the strong coupling modifies the saturation scale, so that it evolves more slowly with the rapidity $Y_0$ as~\cite{munier.peschanski.2004a}
\begin{equation}
	Q_{s,rc}^2(Y_0) = \Lambda_{QCD}^2\exp\left[\alpha_{c} (Y_0+\delta_1)^{1/2}+\beta_c (Y_0+\delta_2)^{1/6}\right], 
	\label{eq:qs_rc}
\end{equation}
where 
\begin{equation}
	\alpha_c = \sqrt{\frac{8\chi'(\gamma_0)}{3}}\simeq 3.61,\quad {\rm and }\quad \beta_c = \frac{3}{4}\xi_1\left(\frac{\chi''(\gamma_0)}{\sqrt{1.5\gamma_0\chi(\gamma_0)}}\right)^{1/3} \simeq -5.36.
	\label{eq:rc_const}
\end{equation}
[$\xi_1=-2.338\hdots$ is the rightmost zero of the Airy function ${\rm Ai}(x)$]. In \cref{eq:qs_rc}, we add correction terms $\delta_{1,2}$ of order $1$, which would be important at finite rapidities. Similar to the fixed coupling case, the largest dipole size in the typical evolution configuration of an onium of size $r$ is also expected to evolve with rapidity as $Q_{s,rc}$,
\begin{equation}
	R_{{s,rc}}^2 (\tilde{Y}_0) = r^2\exp\left[\alpha_c\tilde{Y}_0^{1/2}+\beta_c\tilde{Y}_0^{1/6}\right],
	\label{eq:rs_rc}
\end{equation}
where subleading corrections similar to $\delta_{1,2}$ in~\cref{eq:qs_rc} are neglected. \cref{eq:qs_rc,eq:rs_rc} are up to multiplicative constants. Taking the onium in the dilute regime, it is possible that at a certain stage of the evolution, the typical configuration of the onium at $\tilde{Y}_0$ overlaps with the saturation region of the nucleus evolved to the rapidity $Y_0$, namely $\left[Q_{s,rc}^2(Y_0)R_{s,rc}^2(\tilde{Y}_0)\right]/4 > 1$. Denote $\tau_{0}$ as the smallest scaling variable such that $R_{s,rc}(\tilde{Y}_0)<2/Q_{s,rc}(Y_0)$, for all $Y_0,\tilde{Y}_0 > 0$. We consider two following cases:
\begin{enumerate}[label=(\roman*)]
	\item $\tau>\tau_0$ (see~\cref{fig:theory} (left) for $\tau=0.6, 1.5$). In this case, we can employ the model of large-dipole fluctuations. Following Ref.~\cite{munier.mueller.2018a} and~\cref{subsec:partonic_model}, the rapidity gap distribution is proportional to $P(r,\tilde{Y}_0|2/Q_{s,rc}(Y_0))$. Since $P$ now solves the running coupling BK equation, we can employ its asymptotic solution derived in Refs.~\cite{mueller.trianta.2002,munier.peschanski.2004a}. The asymptotic rapidity gap distribution reads
	\begin{equation}
		\mathcal{R}^{dip} \sim \tilde{Y}_0^{1/6}\left[\frac{Q_{s,rc}^2(Y_0)R_{s,rc}^2(\tilde{Y}_0)}{4}\right]^{\gamma_0}{\rm Ai}\left(\xi_1+\frac{3\xi_1}{4\beta_c}\frac{\ln \left[\frac{4}{Q_{s,rc}^2(Y_0)R_{s,rc}^2(\tilde{Y}_0)}\right]}{\tilde{Y}_0^{1/6}}\right)
		\label{eq:Rdip_rc_theory}
	\end{equation}
	\item $\tau<\tau_0$ (see~\cref{fig:theory} (left) for $\tau=0.3$). For such chosen values of the onium size, it is possible that $R_{s,rc}(\tilde{Y}_0)>2/Q_{s,rc}(Y_0)$ for moderate values of $Y_0$, and $R_{s,rc}(\tilde{Y}_0)<2/Q_{s,rc}(Y_0)$ for $Y_0$ close to $0$ or $Y$. For the latter, \cref{eq:Rdip_rc_theory} can be applied, while for the former, the model of large-dipole fluctuations cannot be employed. On the other hand, as discussed in the previous section, when saturation effects are more significant, the contribution of diffraction at a fixed rapidity gap is less. Therefore, distribution is more suppressed when the largest dipole size of the typical onium's evolution configuration $R_{s,rc}(\tilde{Y}_0)$ becomes larger (and larger than the inverse saturation scale $2/Q_{s,rc}(Y_0)$).
\end{enumerate}

We plot \cref{eq:Rdip_rc_theory} at $Y=30$ in \cref{fig:theory} (right) for different onium sizes and with a particular choice of $\delta_{1,2}$ such that $\delta_1>\delta_2$. Due to the argument in the point (ii) for $\tau<\tau_0$ and the analytical continuity, the distribution for $\tau=0.8$ in the range of $Y_0$ such that $R_{s,rc}(\tilde{Y}_0)>2/Q_{s,rc}(Y_0)$ (the gap in the distribution for $\tau=0.8$) has the hammock shape and the overall distribution is similar to the one for $\tau=1.0$.  When $\tau$ is close to $\tau_0$, the distribution (\ref{eq:Rdip_rc_theory}) is dominated by the Airy function, and large-gap diffraction is favoured. The contribution of the small-gap region becomes more important at smaller onium sizes. At a certain point, the power term $(\dots)^{\gamma_0}$ plays the dominant role, and the large-gap contribution is suppressed. Physically, this suppression is due to the fact that very small onia evolve slowly, so they need a considerably large rapidity interval to develop and create large-dipole fluctuations. Such general behaviors are exhibited in the results presented above. Certainly, our discussion here is, in principle, valid for large $Y,Y_0$ and $\tilde{Y}_0$. Finite-rapidity effects should deform the shape of the distributions predicted asymptotically. 

\subsection{Comparisons to other studies}

Very recently, the diffractive dissociation has been studied in Ref.~\cite{Bendova.etal.2020}, which predicted that about $20\%$ of the events will be diffractive in ePb collisions at $Q^2 = 2\rm\ GeV^2$, and this ratio does not vary much at different values of the Bjorken $x$ (or correspondingly the rapidity $Y$), based on several models. In the current analysis, that ratio is estimated to be about $18\%$ (large-gap threshold) to $27\%$ (small-gap threshold) at $Y=6$, and  about $25\%$ to $35\%$ at $Y=10$ (considering only the fixed-coupling case), which are, in general, fairly higher than the value of $20\%$. In fact, the models present in Ref.~\cite{Bendova.etal.2020} would be applied more appropriately for the diffractive scattering with large rapidity gaps (or high mass). In such regime, our prediction is comparable to the predicted value in the mentioned reference.

In Ref.~\cite{levin.lublinsky.2002a}, the diffractive cross sections at fixed (and large) rapidity gaps in the fixed coupling scenario were plotted for $Y=10$ at different onium sizes close to the saturation scale. They have a similar trend to the case of $Y=10$ in our analysis, with no maximum observed. From our results, the maxium appears only at a large value of rapidity ($Y=30$) (see~\cref{fig:dsigma_dp_30}, in the saturation region). As such maximum was suggested to be the signature of the scaling phenomena, we believe that $N_D$ would exhibit well the scaling behavior at high rapidities.

The mass spectra from our calculation have a quite similar shape to the curves of the same quantity shown in Ref.~\cite{EIC.2016} calculated based on the models of saturation~\cite{Kowalski.etal.2008a,Kowalski.etal.2008b,Toll.Ullrich.2013}, and of leading-twist shadowing (LTS)~\cite{Frankfurt.etal.2004,Frankfurt.etal.2012}. In comparison to the results of the former, there are two different features. Firstly, in the mass spectra shown in Ref.~\cite{EIC.2016} derived from a model of saturation, there is maximum in the low-mass region. Such maximum does not appear in our results. Furthermore, the saturation model predicted a lower-lying distribution at a higher $Q^2$ for all possible values of $M_X^2$. However, our results show that, the $Q^2$-dependence of the mass spectra in the high-mass region is different from the one in the low-mass region. 

Finally, one can make a comment on the nuclear dependence of the distribution of rapidity gaps. Its suppression when increasing $A$ would lead also to the suppression of the diffractive mass spectrum. Such behavior seems to qualitatively agree with the results derived from the LTS model~\cite{EIC.2016,Frankfurt.etal.2004,Frankfurt.etal.2012}, which implies the nuclear shadowing effect. Indeed, the nuclear shadowing is usually explained by multiple scattering (see Ref.~\cite{Armesto.2006} and references therein), which becomes important for large onium sizes ($r\gtrsim 2/Q_s$), i.e. in the saturation region, which is better probed with larger nuclei when fixing the photon virtuality. 

\section{\large CONCLUSION}
\label{sec:conclusion}

To summarize, in the present paper, we have presented numerical results for the distribution of rapidity gaps in the diffractive deep-inelastic virtual photon-nucleus scattering. They are based on the well-established QCD evolution equations at small-$x$ in both fixed and running coupling scenarios. Our main points can be recapped as follows:

\begin{enumerate}[label=(\roman*)]
	\item The diffractive events are predicted to account for a significant fraction in the nuclear scattering of a virtual photon at low to moderate values of the virtuality $Q^2$. At higher $Q^2$, their contribution is suppressed. 
	\item The running coupling correction significantly modifies the rapidity gap distribution in comparison to the case in which the coupling is fixed. Meanwhile, the distribution is relatively independent of the choice of the QCD running coupling prescription.
	\item For the chosen kinematics, large-gap diffraction is more favoured in the deep-inelastic virtual photon-nucleus scattering. The shape of the rapidity gap distributions reflects the shape predicted by the recently developed partonic picture for diffractive dissociation. For the case of fixed coupling, it can be explained completely by the term $-3/(2\gamma_0)\ln Y$ of the saturation scale (see \cref{eq:qs_fc}). Meanwhile, both $Y^{1/2}$ and $Y^{1/6}$ terms in the expression of the running-coupling saturation scale play essential roles in shaping the rapidity gap distribution.
\end{enumerate}

Our analysis again demonstrates that the study of the rapidity gap distribution could reveal the underlying partonic mechanism of diffractive dissociation. Therefore, such observable would be important to be measured at a future electron-ion collider.

The current analysis neglects the impact parameter dependence. While we believe that the general trend is similar when taking into account that dependency, there could be a significant modification on the rapidity gap distribution to be understood. In addition, further developments for the inclusion of subleading corrections appear to be important. Finally, finite-rapidity investigations are of importance for a better understanding of diffractive dissociation in electron-ion collisions.

\section*{\large ACKNOWLEDGEMENTS}

We would like to thank St{\'e}phane Munier for stimulating discussions and comments. This work is supported in part by the Agence Nationale de la Recherche under the project ANR-16-CE31-0019.

\appendix

\section*{\large APPENDIX: NUMERICAL SETUP}
\label{sec:appendix}

The dipole profiles $N$ and $N_{in}$ obey the BK equations in both fixed-coupling and running-coupling scenarios. To solve such integrodifferential equations, we use the fourth-order Runge-Kutta method with a rapidity step size $h_Y = 10^{-2}$. Solutions are stored in a grid of the onium size variable $r$ in which 1000 points are spaced equally in logarithmic scale in the range $10^{-14} \le r\Lambda_{QCD} \le 10^{2}$. Integrals are computed using the mid-point quadrature scheme. For $r\Lambda_{QCD}<10^{-14}$, we use the power-law extrapolation (assuming a power law $\alpha r^\beta$ in the tail), while in the case of $r\Lambda_{QCD}>10^{2}$, we stabilize the profiles at $1$ (saturation). 

Different parameters for the calculation are set as follows:

\begin{enumerate}[label=\roman*.]
	\item The QCD parameter $\Lambda_{QCD}=0.217\ \rm GeV$. This value is obtained by requiring that the value of the running coupling at the mass of the $Z^0$ boson is $\bar{\alpha}_s\left(r^2 = 4C^2/M_{Z^0}^2\right) = 0.1104$~\cite{PDG}, with $M_{Z^0}=91.18 \rm\ GeV$.
	\item Fixed coupling $\bar{\alpha}_s = 0.14$.
	\item The freezing value for the running coupling $\bar{\alpha}_{thres} = 0.5$.
	\item The constant $C$ in the running coupling is manually set to the value $C^2=6.5$ \cite{albacete.etal.2009}~\footnote{In Ref.~\cite{albacete.etal.2009}, the authors used the solution to the running coupling BK equations to fit the data. For the MV initial condition with an anomalous dimension $\gamma$, the fitting value of $C^2$ is $6.5$. Here we take this value as a reference.}. 
	\item Nuclear saturation scale $Q_A^2 = 0.26A^{1/3}Q_{p0}^2$, where $A$ is the nuclear mass number and the saturation scale of proton $Q_{p0}$ at zero rapidity is assumed to be $\Lambda_{QCD}$ . The factor $0.26$ leads to the smallness of the $Q_A^2/Q_{p0}^2$, which was intepreted as a weak nuclear enhancement~\cite{Kowalski.etal.2008b}. 
	\item Masses of quarks $m_u=m_d=m_s=140\ \rm MeV$, $m_c = 1.5\ \rm GeV$. The number of active quarks in the flavor sum in \cref{eq:wave_func_L,eq:wave_func_T} is determined from the condition $Q^2>4m_f^2$.
\end{enumerate}

To check the validation of the numerical calculation, we plot the saturation scales as functions of rapidity and fit them for different scenarios (for $A=208$) (see~\cref{fig:sat_scale}). For the fixed coupling case, the saturation curve is fitted with the following function:
\begin{equation}
	Q_s^{fc} (Y) = a_f\exp\left(b_fY-c_f\ln Y\right).
	\label{eq:sat_scale_fixed}
\end{equation}
Otherwise, the function
\begin{equation}
	Q_s^{rc} (Y) = a_r\exp\left[b_r(Y+d_{r1})^{1/2}+c_r(Y+d_{r2})^{1/6}\right]
	\label{eq:sat_scale_rc}
\end{equation}
is fitted to the data points.

Fitting parameters are shown in \cref{tab:sat_scale_fit}. The fitting values of the leading coefficients $b_f$ and $b_r$ are close to their well-established theoretical values~\cite{mueller.trianta.2002, munier.peschanski.2004a, gribov.etal.1983}, which reads $b_f^{(t)} =\bar{\alpha}_s\chi'(\gamma_0)/2\approx 0.342$ and $b_r^{(t)} =\alpha_c/2\approx 1.804$, respectively. Furthermore, for the fixed coupling case, the value of the next-to-leading coefficient $c_f$ from the fit also approximates to its theoretical value $c_f^{(t)}=3/(4\gamma_0)\approx 1.195$.

\begin{minipage}[t][6.04cm][b]{.44\textwidth}
	\begin{Figure}
		\centering 
		\includegraphics[width=\linewidth]{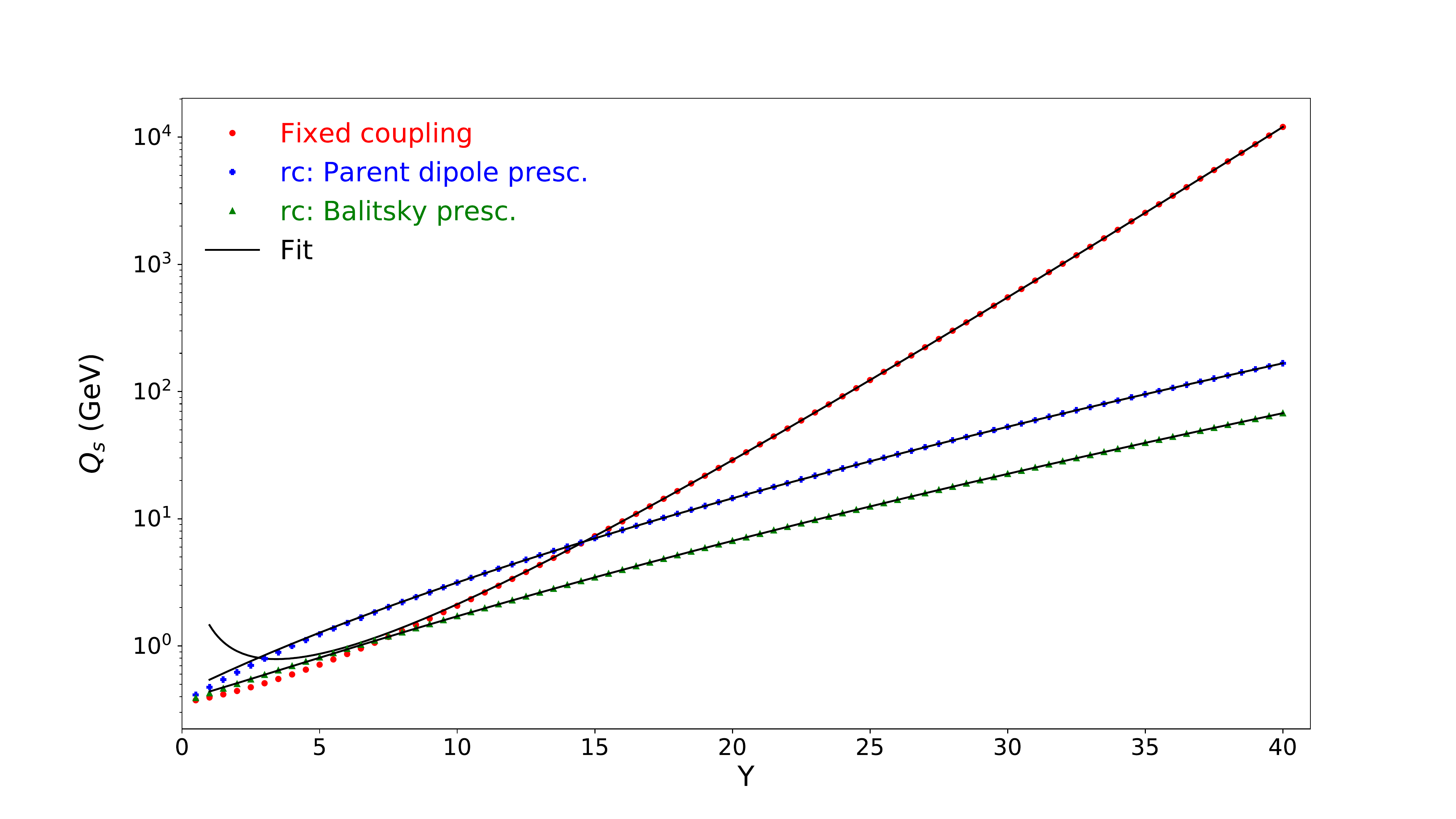}%
		\captionof{figure}{Saturation momenta extracted from the solutions to the BK equations in three different schemes. Black lines are fitting results using functions \cref{eq:sat_scale_fixed} and \cref{eq:sat_scale_rc} (for $Y\ge 1$).}
		\label{fig:sat_scale}
	\end{Figure}
\end{minipage}%
\hfill
\begin{minipage}[t][5cm][b]{.50\textwidth}
	\begin{center}
	\begin{tabular}{|c|ccccc|} \hline
   	Kernel & $a_f$ & $\mathbf{b_f}$ & $c_f$ & & \\ \hline
  	$K^{LO}$ & 1.035 & 0.342 & 1.174 & & \\ \hline\hline
  	Kernel & $a_r$ & $\mathbf{b_r}$ & $c_r$ & $d_{r1}$ & $d_{r2}$ \\ \hline
  	$K^{pd}$ & 0.419 & 1.805 & -3.374 & 7.862 & 11.131 \\ \hline
  	$K^{Bal}$ & 0.111 & 1.810 & -3.352 & 9.432 & 4.635 \\ \hline
  	\end{tabular}
  	\vspace{1.18em}
  	\captionof{table}{Values of the parameters in \cref{eq:sat_scale_fixed,eq:sat_scale_rc} obtained from corresponding fits to the numerical data points shown in \cref{fig:sat_scale}.}
  	\label{tab:sat_scale_fit}
	\end{center}
\end{minipage}

\renewcommand{\refname}{REFERENCES}
\bibliographystyle{apsrev4-1}
\small
\bibliography{DDIS.bib}
\end{document}